\documentclass{article}
\usepackage{ijcai22}

\usepackage{hyperref}
\usepackage{url}
\usepackage{algorithm}
\usepackage{algpseudocode}
\usepackage{graphicx}
\usepackage{amsmath}
\usepackage{amssymb}
\usepackage{subcaption}

\title{The Good Shepherd: An Oracle Agent for Mechanism Design}

\author{Jan Balaguer, Raphael K\"oster, Christopher Summerfield,\And Andrea Tacchetti
\affiliations
DeepMind, UK
\emails
\texttt{\{jua, rkoster, csummerfield, atacchet\}@deepmind.com}}

\begin{document}

\maketitle

\begin{abstract}

    From social networks to traffic routing, artificial learning agents are playing a central role in modern institutions. We must therefore understand how to leverage these systems to foster outcomes and behaviors that align with our own values and aspirations. While multiagent learning has received considerable attention in recent years, artificial agents have been primarily evaluated when interacting with fixed, non-learning co-players. While this evaluation scheme has merit, it fails to capture the dynamics faced by institutions that must deal with adaptive and continually learning constituents. Here we address this limitation, and construct agents (``mechanisms'') that perform well when evaluated over the learning trajectory of their adaptive co-players (``participants'').
    The algorithm we propose consists of two nested learning loops: an inner loop where participants learn to best respond to fixed mechanisms; and an outer loop where the mechanism agent updates its policy based on experience.
    We report the performance of our mechanism agents when paired with both artificial learning agents and humans as co-players. Our results show that our mechanisms are able to shepherd the participants strategies towards favorable outcomes, indicating a path for modern institutions to effectively and automatically influence the strategies and behaviors of their constituents.
\end{abstract}

\section{Introduction}
\begin{figure*}[ht]
    \centering
    \includegraphics[width=\textwidth]{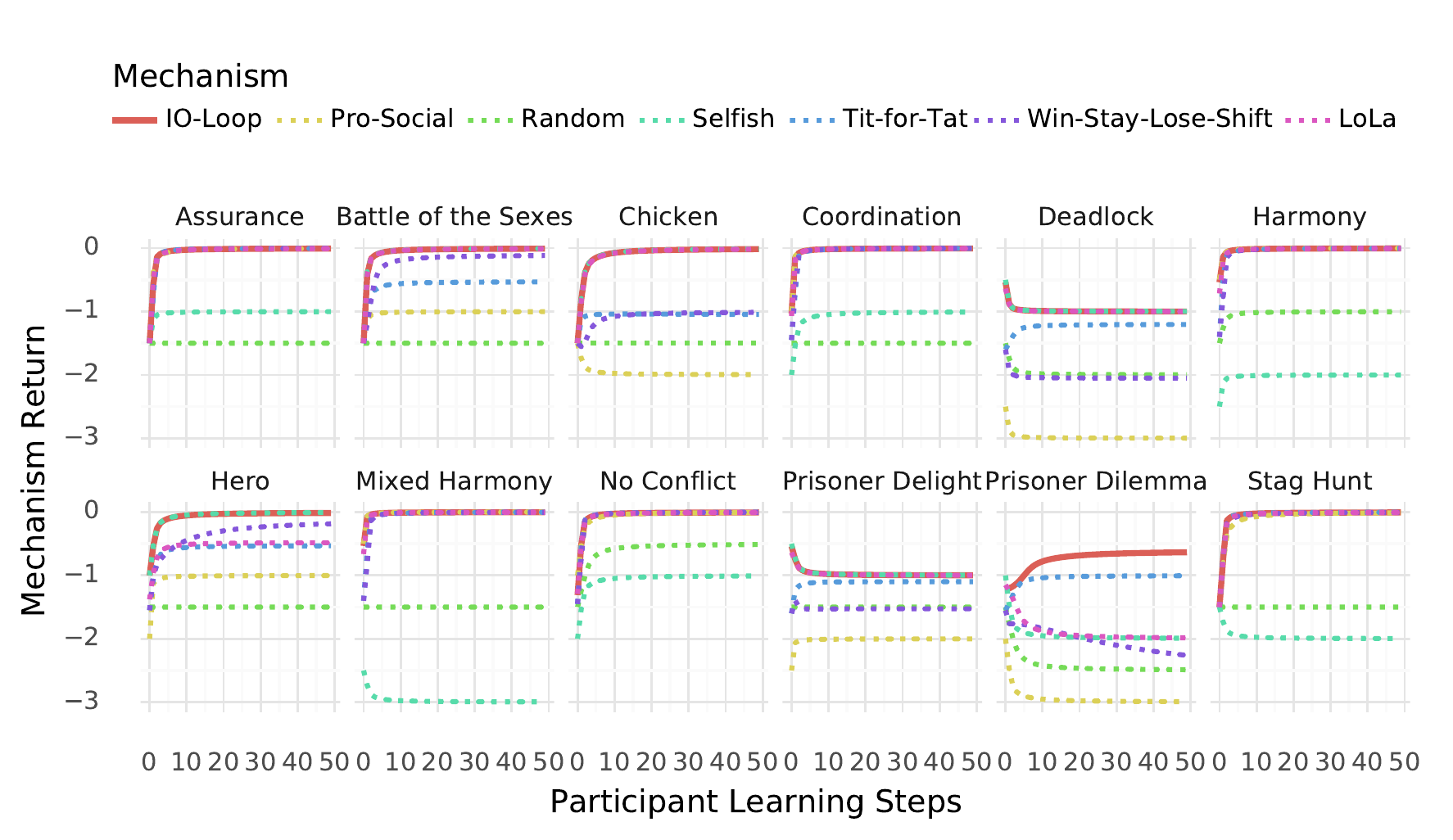}
    \caption{\small Performance of fixed mechanisms paired with adaptive participants in the 12 matrix games we consider. The horizontal axis shows the learning steps of the participant agent, while the vertical axis shows the return collected by the fixed mechanism agent. Our IO-loop agent, here trained with Diff-MD, tracks the best alternative strategy in all games but Prisoner Dilemma where it significantly outperforms it. Plots are produced averaging mechanism returns over 5 random seeds.}
    \label{fig:diff-md}
\end{figure*}    
Modern institutions often serve two distinct and equally important roles in our society. First they mediate and foster economic or social interactions among citizens (e.g. taxation policies ensure governments receive enough funds to build roads and schools). Second, they foster behaviors that bring us closer to our aspirations as a society (e.g. charitable donations are tax-free). As artificial learning agents mediate more and more interactions among humans, firms, and organization, it is paramount that we study how to construct adaptive systems that can fulfil both roles.

However, while multiagent learning has received considerable attention in recent years, the standard evaluation scheme pairs our artificial agents with other fixed, and potentially adversarial co-players (e.g. exploitability)~\shortcite{vinyals2019grandmaster,muller2019generalized,goodfellow2014generative}. While this evaluation scheme has merit, it fails to capture the dynamics faced by modern institutions that are often paired with learning constituents, and where agents must take into account not only what other agents will do next, but also, in the long run, how they will adapt to the current strategies present in the system.

Here we address this shortcoming and construct low-exploitability agents that do well when paired with learning co-players, in the general-sum setting. We construct players that, through their behavior, are able to influence what others will learn to do, and explicitly leverage the link between one agent's actions and another agent learning trajectory. In other words, we construct agents (``mechanisms'') that learn to act so as to shepherd participants' strategies both at equilibrium, and during learning.

Our proposed method takes the form of an inner-outer loop learning process. In the inner loop, participant agents respond to a fixed mechanism strategy, while in the outer-loop our mechanism agent adapts its policy based on experience. Unlike previous work, our mechanisms make very few assumptions on the preferred strategies, outcomes, or learning capabilities of the participants, and only have access to the consequences of their own behavior on the learning of others.

We investigate the performance of our mechanism agents with both artificial and human co-players in simple 2-player 2-strategy repeated games, and in a stylized resource allocation problem. We study how our method can be adapted when mechanisms are granted access to the inner workings of participants, and when that is not the case. 

Our results show that our mechanism agents successfully shepherd the learning of others towards desirable outcomes, and that the direction presented here is promising for agent-agent interactions, and withstand a transfer to the agent-human interaction setting.

In the broader context of AI in modern day institutions, our methods and ideas show that adaptive agents can successfully shepherd the learning of their co-players towards desirable outcomes and behavior, opening the door to learning-based institutions that fine-tune the incentives faced by their constituents in pursuit of group level goals.

 \section{Related work}
\begin{figure*}[ht]
    \centering
    \includegraphics[width=\textwidth]{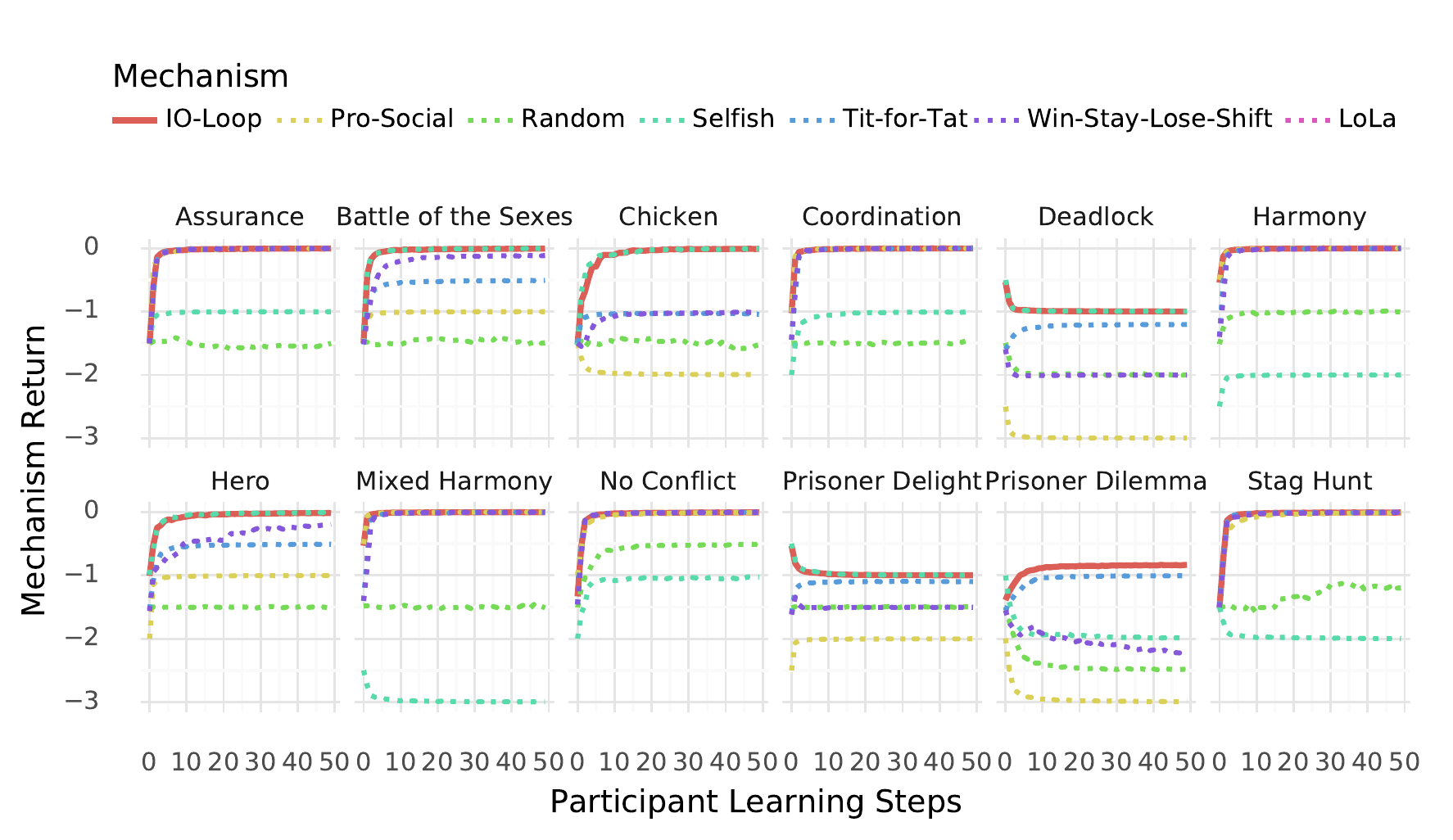}
    \caption{\small Performance of fixed mechanisms paired with adaptive participants in the 12 matrix games we consider. The horizontal axis shows the learning steps of the participant agent, while the vertical axis shows the return collected by the fixed mechanism agent. Our IO-loop agent, here trained with ES-MD, tracks the best alternative strategy in all games but Prisoner Dilemma where it significantly outperforms it. Plots are produced averaging mechanism returns over 5 random seeds.}
    \label{fig:es-md}
\end{figure*}
The inner-outer loop method that we propose here provides insights into two challenges for multiagent reinforcement learning.

The first challenge is the non-stationarity of the environment. When training multiple policies simultaneously, the environment is non-stationary from the point of view of any agent due to the change in the other players' policies. Common approaches to mitigate this non-stationarity involve building populations of agents~\shortcite{brown1951iterative,muller2019generalized,vinyals2019grandmaster} or exploiting knowledge of the learning dynamics of others~\shortcite{balduzzi2018mechanics,hemmat2020lead}. Both these approaches have often focused on competitive zero-sum game.
Here we focus on setting where the environment is always stationary from the point of view of all agents since we don't update policies concurrently in the same training loop.

Second is the challenge of \textit{equilibrium selection}. Generally, and particularly in non-zero-sum games, multiple Nash Equilibria may exist. This leads to the problem of both finding and selecting among possibly unequal equilibria, with the goal of a) biasing learning towards outcomes preferred by one agent (shaping), or b) generalizing with unseen co-players. Progress towards this has been made in recent years through \textit{centralized learning with decentralized execution}, a framework for multiagent RL where agents can exploit privileged knowledge about other agents' during training, but not at time of deployment. Centralized learning can be useful for equilibrium selection: for example, access to a centralized value function provides a recipe to construct agents that are able to coordinate at execution in cooperative settings~\shortcite{sunehag2017value}; coupled training of multiple agents can improve learning of communication protocols~\shortcite{foerster2016learning,foerster2019bayesian}; learning from interactions with agents at different stages of training can improve generalization at evaluation with human participants~\shortcite{strouse2022collaborating}; and exploiting information about how other agents update their behaviour can be used to shape them, both within an episode~\shortcite{lerer2018maintaining,peysakhovich2018consequentialist} and across training~\shortcite{foerster2017learning,yang2020learning}.  
In our method, we do not separate training from execution. We make no assumptions about the learning rule of the co-players, or how they will adapt to the strategies currently present in the system. Instead, we infer the relationship between the mechanism's actions and the participant's learning directly from the observed interactions.

 \section{Methods}
 We consider the problem of constructing agents (``mechanisms'') that shepherd the learning of others (``participants''). We use repeated symmetric two-player, two-strategies games as our initial test-bed as they are easy to analyze and train on. We then move on to a simple resource allocation game. We start by assuming that the mechanism agent has access to the inner workings of participant agents in Differentiable mechanism design, and then extend our methods to remove this assumption using Evolutionary Strategies~\shortcite{salimans2017evolution}.
     
    \subsection{Environments}
    We tackle iterated 2-player, 2-strategies symmetric matrix games, and a simple resource allocation game with one mechanism agent and four participant agents.

        \subsubsection{Iterated matrix games}
        \begin{figure*}[ht]
            \centering
            \captionsetup[subfigure]{labelformat=empty}
            \begin{subfigure}{.48\textwidth}
            \centering
            \includegraphics[width=\textwidth]{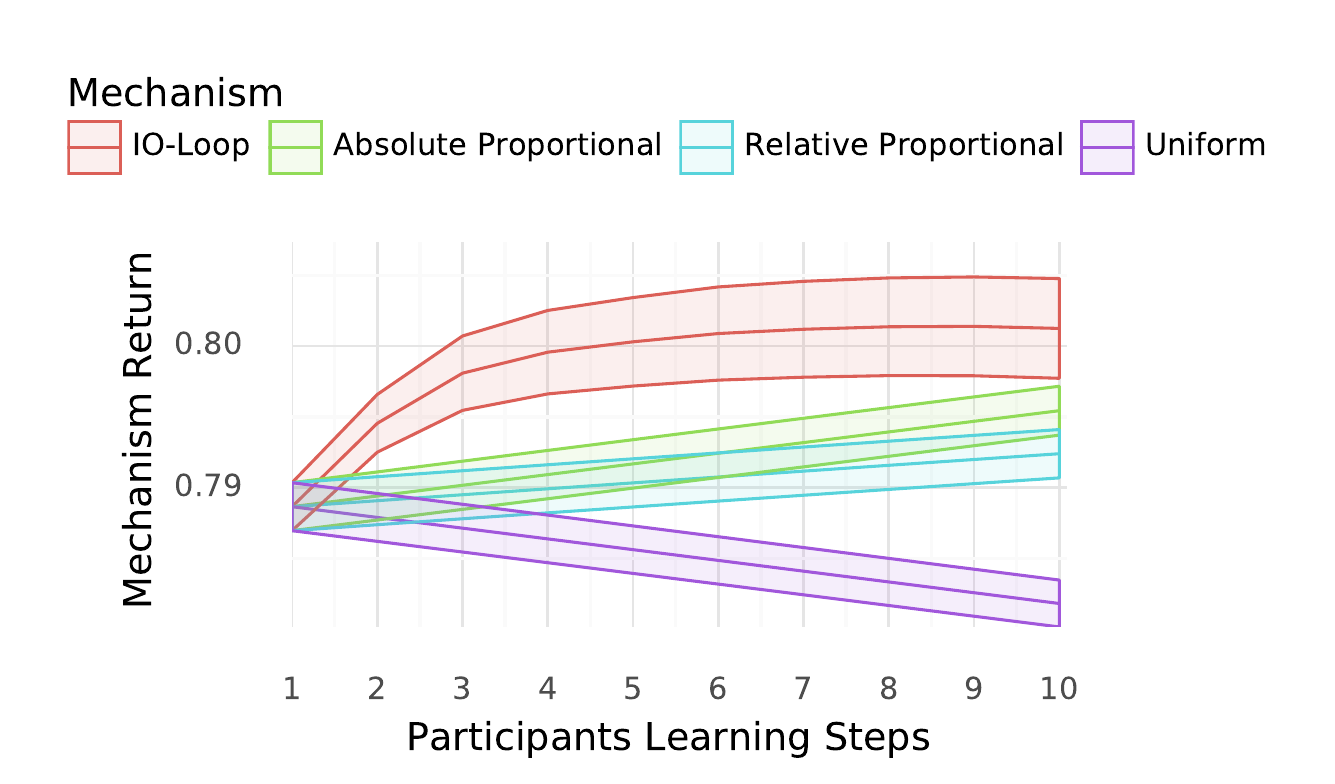}
            \end{subfigure}
            \begin{subfigure}{.48\textwidth}
            \centering
            \includegraphics[width=\textwidth]{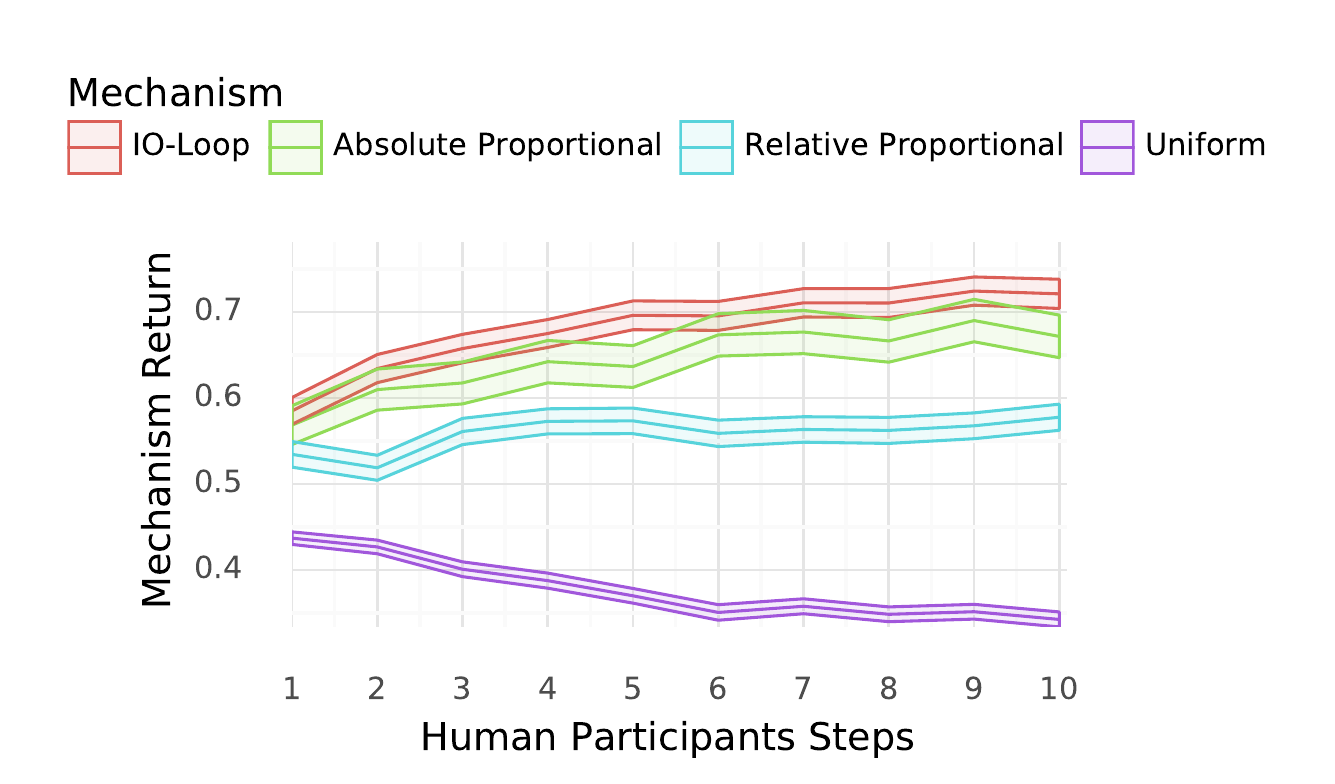}
            \end{subfigure}            
            \caption{\small Performance of fixed mechanisms paired with artificial adaptive participants (left) or human participants (right) in the resource allocation game we consider. The horizontal axis shows the learning steps of the participant agents (independent learning), while the vertical axis shows the return collected by the fixed mechanism agent. Our IO-loop agent, here trained with Diff-MD, outperforms alternative mechanisms proposed in the economics literature for this game both with artificial learning and human co-players. The left panel was produced averaging mechanism returns over 50 random seeds, while the right panel shows the average return over each experiment; shaded areas indicate standard error.}
            \label{fig:pgg}
        \end{figure*}        
        We consider the 12 symmetric 2-player, 2-strategies matrix games identified in~\shortcite{wiki:Normal-form_game,robinson2005topology}, with payouts scaled down by 4 (our payouts are between -3 and 0), and consider iterated interactions between a mechanism (row) player and a participant (column) player with a single memory step. Our naming convention is most easily followed when focusing on the Prisoner Dilemma game.
        
        For each game we define a Markov Decision Process with statespace $\mathcal{S} = (s_0, CC, CD, DC, DD)$, with $s_0$ being the initial state, and the rest being the state after the joint actions in the previous state (e.g. cooperate, cooperate; cooperate, defect and so on). Our agents' one-memory policies can be represented by a 5-tuple $\theta$ corresponding to the probability of cooperating on each state. In these simple repeated games, the transition kernel $\mathcal{T}$ takes the form of a matrix whose entries describe the probability of the next state as a function of the previous state, and can be derived analytically given the one-memory policy parameters from both players. The reward functions are specified on arrival at each state as $r_m = (0, r_{R}, r_{S}, r_{T}, r_{P})$ and $r_p =(0, r_{R}, r_{T}, r_{S}, r_{P})$, where $r_{P}$, $r_{R}$, $r_{S}$, $r_{T}$ correspond to the punishment, reward, sucker and temptation payoffs respectively. The returns $R_m$ and $R_p$ that both the mechanism and participant aim to maximize corresponds to the state value for the initial state $s_0$.
        
        \subsubsection{Resource Allocation game}
        We further consider a modification of the classic Public Goods Game (as described in~\shortcite{koster2022humancentered}). The game consists of a single interaction between four participants $i = 1,2,3,4$ and a single mechanism agent. Each participant receives an endowment $e_i$ and allocates a fraction $\rho_i$ of it to a common investment pool, which is then grown by a fixed constant factor ($1.6$) and redistributed to participants in full. The specific amount received by each participant $i$ is denoted as $p_i$ and is determined by the mechanism. Participants seek to maximize their individual welfare (i.e. $R_{p_i} = p_i + (1-\rho_i)e_i$), while the mechanism seeks to maximize total participants' welfare (i.e. $R_m=\frac{1}{N}\sum_i^N  R_{p_i}^t$, with $N$ the number of participants). When interacting with naive or poorly designed redistribution mechanisms, the incentive to free-ride may tempt each player away from the contributing to the common pool, which in turn decreases total welfare.

    \subsection{The Inner outer loop Algorithm}
    \begin{algorithm}[ht]
        \begin{algorithmic}
            \Require $\mathcal{MDP}, T_m, T_p, \mathfrak{R}, \theta_m^0 ,\gamma_m, \gamma_p$
            \For{$t_{m} \text{ in } 0 :: T_{m}$}
                \State $\theta_p^0 \sim \mathfrak{R};\quad \bar{R}_m \gets 0$
                \For{$t_{p} \text{ in } 0 :: T_{p}$}
                    \State $(R_m, R_p) \gets \mathcal{MDP}(\theta_m^{t_m}, \theta_p^{t_p})$
                    \State $\bar{R}_m \gets \bar{R}_m + R_m$
                    \State $\theta_p^{t_p+1}\gets \theta_p^{t_p} + \gamma_p\nabla_{\theta_p^{t_p}} R_p$
                \EndFor
                \State $\theta_m^{t_m+1} \gets \theta_m^{t_m} + \gamma_m\nabla_{\theta_m^{t_m}} \bar{R}_m$
            \EndFor
        \end{algorithmic}
        \caption{\small Inner outer loop algorithm. Given an underlying Game function ($\mathcal{MDP}$) that takes policy parameters $\theta_m$ and $\theta_p$ for the mechanism and participants respectively computes their returns $R_m$ and $R_p$, a random participant parameter generator $\mathfrak{R}$, and initial mechanism parameters $\theta_m^0$ this algorithm produces a policy $\theta_m^{T_m}$ for the mechanism player.}\label{algo:ioloop}
    \end{algorithm} 

    Our learning process takes the form of an inner-outer loop that exposes the mechanism to the consequences of its actions on the learning of others. In the inner loop, participant agents repeatedly interact with a fixed mechanism, and use independent gradient ascent to improve their own policies. In the outer loop, mechanism agents update their strategies based on the experience they acquired in the inner loop. 
    
        \subsubsection{Differentiable mechanism design}
        We first consider the case where the mechanism agents can directly compute the gradients of its return $\bar{R}_m$ with respect to its policy parameters $\theta_m$. Inspecting Algorithm~\ref{algo:ioloop} we note that, at any given $\theta_m^{t_m}$ update, the mechanism return $R_m$ depends on the mechanism parameters $\theta_m^{t_m}$, as well as on the entire trajectory of participants parameters over the inner loop $\theta_{p_0},\ldots,\theta_p^{T_p}$. In Differentiable Mechanism design (Diff-MD), we let the mechanism update have gradient access to the entire trajectory, as well as the environment transition kernel $\mathcal{T}$.
        In practice, this can easily be implemented using a tensor library with auto-differentiation (such as JAX~\shortcite{jax2018github}).

        \subsubsection{Evolutionary strategies for non-differentiable mechanism design}

        When the mechanism agent cannot take derivatives through the environment, we used evolutionary strategies (ES). In this case, the inner loop is repeated $N_p$ times so as to form an experience ``batch'' with mechanism parameters slightly perturbed at the beginning of each inner loop ($t_p=0$) as $\theta_{m}^p = \theta_{m}^p + \epsilon_p$, where $\epsilon_p \sim \mathcal{N}(0,\sigma_{m}^{2})$ with $\sigma_m$ a hyper-parameter ($1$ in our experiments). Given an experience batch the mechanism policy gradient, estimated as $\nabla_{\theta_{m}} \approx \sum_{p=1}^{N_p} \frac{\epsilon_{p} \bar{R}_m}{N_{p} \sigma_{m}}$, moves the mechanism parameters in the direction of those used in episodes that led to positive outcomes (see~\shortcite{salimans2017evolution} for details). We refer to this method as ES-MD.

    \subsection{Learning with Opponent-Learning Awareness (LOLA)}

    We implemented LOLA~\shortcite{foerster2017learning} as a baseline in our experiments. In LOLA, the mechanism agent projects the learning of the participants forward in time. In contrast to the original paper, in which both agents are assumed to be using LOLA, here only let the mechanism agent be learning-aware.

\section{Results}
Here we show how a trained mechanism performs when paired with learning participants. We report the return collected by a (fixed) mechanism in each episode over the learning trajectory of its co-players. Figures~\ref{fig:diff-md} and~\ref{fig:pgg} show how mechanisms trained with Diff-MD perform in the 12 matrix games and resource allocation game respectively, while Figure~\ref{fig:es-md} shows the performance of a mechanism trained with ES-MD in the 12 matrix games we consider. 

\subsection{Matrix games}

In the matrix games we compare our mechanism (labelled as IO-Loop in the legends) to well known one memory strategies (e.g. Tit-for-Tat) and pure strategies (e.g. Selfish). Additionally, in Diff-MD we also compare a mechanism trained with LOLA. In all games, and for both Diff-MD and ES-MD, our IO-loop mechanism achieves the same performance as the best alternative available strategy, with the exception of Prisoner Dilemma where it significantly outperforms it. We used the following hyper-parameters for both experiments: $T_m=10000, T_p=50, \gamma_m=0.1, \gamma_p=10,\theta_m^0=[0.5, 0.5, 0.5, 0.5, 0.5]$, in the ES-MD experiment we further set $N_p=256$ and $\sigma_m=1$.

\subsection{Resource allocation game}
In the resource allocation game we report mechanism performance when paired both with artificial learning agents and human co-players. In particular, in Fig.~\ref{fig:pgg}, we consider unequal endowments with $e_i \in [0.2,1.0]$, and compare our mechanism with four alternative redistribution strategies: Absolute Proportional and Relative Proportional redistribute funds proportionally to the absolute contribution $\rho_i e_i$ or to the fraction of endowment $\rho_i$ contributed by each participants, while the Uniform and Random mechanisms redistributed the funds equally and randomly respectively. Figure~\ref{fig:pgg} shows that Diff-MD finds a mechanism policy that shepherds the participants towards higher welfare outcomes. In this experiment we represented the participants policy as their propensity to contribute to the public fund:  $\theta_{p_i}=\rho_i$, and the mechanism policy as a MLP with a single 32-units hidden layer. We further set $T_m=5000, T_p=10, \gamma_m=0.01, \gamma_p=0.1$ and $\theta_m^0$ the default MLP initialization.

\subsection{Evaluation with human co-players}

Figure~\ref{fig:pgg} (right) shows the performance of our mechanism, and alternative baselines, when paired with human co-players in the stylized resource allocation game outlined above (endowment condition for the 4 players: $[1.0, 0.5, 0.4, 0.3]$). We used crowd-sourcing platforms to collect data, and all participants gave informed consent to participate in the experiment. Participants were organized in groups of 4, and after a tutorial phase, they played the resource allocation game outlined above completing the 10 steps constituting our ``inner loop''. The tutorial round explained the mechanics of the game, instructed participants on how to use the web interface, and outlined how participants would be rewarded in real money: participants received a base compensation for completing the experiment, and a bonus proportional to their aggregate return over the course of the experiment. After one game with the Uniform mechanism, each group of participants interacted with two mechanisms, either resulting from our training, or with one of the baseline mechanisms outlined above in counterbalanced order. We collected data from 236 non overlapping groups. If a participant dropped out during the experiment, their actions were replaced with random actions, which were subsequently removed in the analysis ($39\%$ of responses were removed this way). The results presented in the right hand panel in Figure~\ref{fig:pgg} show that our mechanism could withstand a basic transfer to interacting with human co-players, and that its performance remained consistent with what we observed in simulation.

\section{Conclusion}
We have shown here that our inner-outer loop algorithm can provide an oracle-style benchmark to test agents' ability to shepherd the behavior of learning co-players to desired outcomes, and that in simple environments, our agents transfer to human co-players.

As more and more of the systems we use and deploy become adaptive, it becomes increasingly important to 1) construct agents that can plan and act taking into account the fact that others are learning around them, and 2) construct agents that can shape the incentives faced by co-players in pursuit of group-wide objectives.

The ideas and results presented here show that exposing agents to the consequences of their actions on the learning of others is a sensible first step toward these goals. Moreover, the transfer to human co-players we were able to showcase suggests that our method contains the basic elements required to design adaptive institutions can fulfill their basic ``mechanical'' mediation function in society, as well as shepherd their constituents towards more desirable strategies and behaviors.

\newpage

\bibliographystyle{plain}
\bibliography{main}

\end{document}